\documentclass[[12pt,a4paper,preprint,preprintnumbers,amsmath,amssymb,nofootinbib]{revtex4}
\usepackage{graphicx}
\usepackage{amsmath}
\usepackage{amssymb}
\usepackage{bbold}
\usepackage{type1cm}
\usepackage{rotating}
\usepackage{tabularx}
\usepackage{pdflscape}

\usepackage{dsfont}
\usepackage{overpic}
\usepackage{natbib}
\usepackage{slashed}
\usepackage{subfigure}
\usepackage{bbm}

\begin{document}
\title{Structure of resonances in a simple quantum-mechanical model}

\author{Peter C.~Bruns}
\affiliation{Nuclear Physics Institute, 25068 \v{R}e\v{z}, Czech Republic }
\date{\today}
\begin{abstract}
We study the structure of resonances derived from the solution of an exactly solvable Lippmann-Schwinger equation. Within this framework, we discuss the concept of ``resonance form factors'', and the description  of the resonant amplitudes in terms of effective energy-dependent potentials. 
\end{abstract}

\maketitle

\section{Introduction}
\label{sec:Intro}

Most hadrons are unstable under the strong interaction, and therefore appear as resonances in various strong-interaction processes, which are described, on a fundamental level, by Quantum Chromodynamics (QCD). For a full understanding of the spectrum and dynamics of QCD, it is necessary to not only determine the partial widths and pole positions of the resonances on the complex energy-surfaces of the reaction amplitudes, but also to get a clue on their structure. Investigations on the structure (or, even more generally, ``nature'') of resonances usually deal with concepts like resonance form factors, ``resonant states'', $K$-matrix poles, energy-dependent potentials and ``compositeness'', all of which can only be extracted from experimental data with some model dependence. It is the aim of the present contribution to explore how meaningful and unambiguous the above-mentioned concepts are, when applied to the solution of an exactly-solvable Lippmann-Schwinger equation (LSE) for a quantum-mechanical potential scattering problem in one space dimension. The potential has been chosen to be as simple as possible, while showing all the interesting features (bound states, resonances, non-trivial momentum dependence -\,but no energy-dependence\,- of the (off-shell) potential (kernel of the LSE)) that we want to investigate. \\
The solution of the LSE is an off-shell scattering amplitude (i.e., the momenta of the scattered particles are not restricted to their mass shells), as pointed out e.g. in \cite{Taylor}. This allows us to extract the wave functions of the bound states of the potential directly from the scattering amplitude (up to a phase), and thus to compute associated form factors which give information on the spatial structure of the state. Of course, the wave functions can also be found directly, by the methods of elementary-school quantum mechanics (we give the explicit expressions in App.~\ref{app:LSdoubledeltawf}); this provides a convenient check. Employing an analytic-continuation argument, we will try to extend the concept of form factors also to the resonances; the inverse Fourier transforms of the resulting expressions will generate certain spatial ``density functions'' related to the resonances. As will become clear along the way, one would have to study the scattering process in the resonance region, probed by some additional external (e.g. electromagnetic) field, in order to extract those functions. An example for a theoretical study of such a process, for the case of the $\Lambda(1405)$, can be found in \cite{Sekihara:2008qk}. \\
This article is organized as follows: In Sec.~\ref{app:LS1d}, we give an outline of the LSE framework for this work to be self-contained, and to fix the notation and conventions. The model studied here will be introduced in Sec.~\ref{app:LSdoubledelta}. Instead of the simplest-possible example, a delta-function potential $\sim v\delta(x)$ with a constant coupling strength $v$, we adopt the next-simplest example: a potential consisting of two delta functions, separated by a distance $2d$. This already suffices to provide us with many interesting properties of the solution of the corresponding LSE, without the need to introduce an energy-dependence of the coupling strength ``by hand''. The most important part of the present work is Sec.~\ref{app:FF}: there, we show how to compute the form factors of the bound-states and resonances from the off-shell solution of the LSE. The ensuing Sec.~\ref{sec:edepv} is de\-dicated to explore an effective description of (the parity-even part of) the scattering amplitude in terms of an {\em energy-dependent\,} potential, to make contact with similar studies (usually in relavistic quantum mechanics, or quantum field theory) in the literature. Our conclusions can be found in Sec.~\ref{sec:Conclusio}; the appendices contain some additional material and explicit expressions. 


\section{Lippmann-Schwinger equation: Generalities}
\label{app:LS1d}

In this work, we consider non-relativistic quantum mechanics on a one-dimensional space. The Schr\"odinger equations are of the form $\hat{H}|\psi_{n}\rangle = E_{n}|\psi_{n}\rangle\,$, $\hat{H}=\hat{H}_{0}+\hat{V}\,$, $\hat{H}_{0}|p\rangle = E_{n}^{(0)}|p\rangle=\frac{|p|^2}{2\mu}|p\rangle\,$, where $\hat{H}$ is the full Hamilton operator, while $\hat{H}_{0}$ is its ``free'' part. Throughout, we shall use hats over a symbol to indicate its operator character, and admit only local potentials $\hat{V}$, i.e. $\langle y|\hat{V}|x\rangle = V(x)\delta(x-y)\,$. We define resolvent functions (operators) $\hat{G}_{0}(E)=(E-\hat{H}_{0})^{-1}\,,\quad \hat{G}(E)=(E-\hat{H})^{-1}\,$ for the free and the full Hamilton operator, respectively. 
Expressing $\hat{H}$ as a differential operator in coordinate space, $G(x',x):=\langle x'|\hat{G}(E)|x\rangle$ solves
\begin{equation}
(E-\hat{H})G(x',x) = \delta(x'-x) = \sum_{n}\psi_{n}(x')\psi^{\ast}_{n}(x)\,,\qquad \langle x|\psi_{n}\rangle =: \psi_{n}(x)\,,
\end{equation}
employing the completeness relation for an orthonormal set of eigenfunctions of $\hat{H}$ (the summation is to be replaced by integration for the continuous part of the spectrum). In operator form, we can express $\hat{G}$ more generally as
\begin{equation}
\hat{G}(E) = \sum_{n}\frac{|\psi_{n}\rangle\langle\psi_{n}|}{E-E_{n}}\,.\qquad\mathrm{Similarly,}\quad \hat{G}_{0}(E) = \int dp\frac{|p\rangle\langle p|}{E-\frac{p^2}{2\mu}}\,.
\end{equation}
We normalize the eigenstates of $\hat{H}_{0}$ as $\langle p'|p\rangle = \delta(p\,'-p)$, so that
\begin{equation}
\langle x|p\rangle = \frac{e^{ipx}}{\sqrt{2\pi}}\qquad\mathrm{and}\qquad \sqrt{2\pi}\langle p|\psi_{n}\rangle =\int dx \,e^{-ipx}\psi_{n}(x)=:\tilde{\psi}_{n}(p)\,.
\end{equation}
In operator form, the Lippmann-Schwinger equation (LSE) reads (see e.g. \cite{Taylor})
\begin{equation}\label{eq:LSE}
\hat{\mathcal{T}}(E)=\hat{V}+\hat{\mathcal{T}}(E)\hat{G}_{0}(E)\hat{V}\,.
\end{equation}
One can convince oneself that the solution can also be written in terms of the full resolvent $\hat{G}$,
\begin{equation}\label{eq:LSG}
  \hat{\mathcal{T}}(E)=\hat{V}+\hat{V}\hat{G}(E)\hat{V}\,,
\end{equation}
since $\hat{G}=\hat{G}_{0}+\hat{G}\hat{V}\hat{G}_{0}=\hat{G}_{0}+\hat{G}_{0}\hat{V}\hat{G}$. Taking matrix elements of Eq.~(\ref{eq:LSE}), we get
\begin{equation}
\langle q'|\hat{\mathcal{T}}(E)|q\rangle = \langle q'|\hat{V}|q\rangle + \int dl\,\frac{\langle q'|\hat{\mathcal{T}}(E)|l\rangle\langle l|\hat{V}|q\rangle}{E-\frac{l^2}{2\mu}}\,.
\end{equation}
 On the other hand, taking matrix elements of (\ref{eq:LSG}),
\begin{eqnarray}
  \langle q'|\hat{\mathcal{T}}(E)|q\rangle &=& \langle q'|\hat{V}|q\rangle + \sum_{n}\frac{\langle q'|\hat{V}|\psi_{n}\rangle\langle\psi_{n}|\hat{V}|q\rangle}{E-E_{n}}\nonumber \\
  &=& \langle q'|\hat{V}|q\rangle + \sum_{n}\frac{\left(\frac{q'^2}{2\mu}-E_{n}\right)\tilde{\psi}_{n}(q')\tilde{\psi}_{n}^{\ast}(q)\left(\frac{q^2}{2\mu}-E_{n}\right)}{2\pi(E-E_{n})}\,,\label{eq:LSGmat}
\end{eqnarray}
by means of the Schr\"odinger equation $\langle q|\hat{V}|\psi_{n}\rangle=-\langle q|\hat{H}_{0}-E|\psi_{n}\rangle\,$ and the completeness relations. The matrix element $\langle q'|\hat{\mathcal{T}}(E)|q\rangle$ is the off-shell scattering amplitude for incoming and outgoing momentum $q$ and $q'$, respectively, and energy $E$. \\

\underline{Normalization condition for the LSE:} \,Writing out the solution of the LSE, Eq.~(\ref{eq:LSE}), as an infinite series,
\begin{equation}\label{eq:LSEser}
\hat{\mathcal{T}}(E)=\hat{V}+\hat{V}\hat{G}_{0}(E)\hat{V}+\hat{V}\hat{G}_{0}(E)\hat{V}\hat{G}_{0}(E)\hat{V} + \ldots\,,
\end{equation}
it is easily seen that
\begin{eqnarray}
\frac{\partial}{\partial E}\hat{\mathcal{T}}(E) &=& \hat{V}'+\hat{V}'\hat{G}_{0}(E)\hat{\mathcal{T}}(E) + \hat{\mathcal{T}}(E)\hat{G}_{0}(E)\hat{V}' + \hat{\mathcal{T}}(E)\hat{G}_{0}(E)\hat{V}'\hat{G}_{0}(E)\hat{\mathcal{T}}(E) \nonumber \\ &+& \hat{\mathcal{T}}(E)\left(\frac{\partial}{\partial E}\hat{G}_{0}(E)\right)\hat{\mathcal{T}}(E)\,,\quad \hat{V}'\equiv \frac{\partial \hat{V}}{\partial E}\,.\label{eq:LSEserder}
\end{eqnarray}
Now, close to a certain eigenvalue $E_{n}$ in the discrete part of the spectrum, we can write
\begin{equation}
\langle q'|\hat{\mathcal{T}}(E)|q\rangle = \frac{\left(\frac{q'^2}{2\mu}-E_{n}\right)\tilde{\psi}_{n}(q')\tilde{\psi}_{n}^{\ast}(q)\left(\frac{q^2}{2\mu}-E_{n}\right)}{2\pi(E-E_{n})} + \mathcal{R}(E)\,,
\end{equation}
where $\mathcal{R}(E)$ is regular as $E\rightarrow E_{n}$. Inserting this ansatz in (\ref{eq:LSEserder}), and identifying the terms with the double pole on both sides, we find the constraint
\begin{equation}\label{eq:LSEBnorm}
\int\frac{dl}{2\pi}\tilde{\psi}_{n}^{\ast}(l)\tilde{\psi}_{n}(l) - \int\frac{dl'}{2\pi}\int\frac{dl}{2\pi}\tilde{\psi}_{n}^{\ast}(l')2\pi\langle l'|\hat{V}'(E\rightarrow E_{n})|l\rangle\tilde{\psi}_{n}(l) \overset{!}{=} 1\,.
\end{equation}
The second term on the l.h.s. only appears for energy-dependent potentials. We will discuss such a potential shortly in Sec.~\ref{sec:edepv}.\\

\underline{``Loop'' integral and Riemann sheets:} The integral
\begin{equation}\label{eq:LSEloop}
I_{0}(E):=\int_{-\infty}^{\infty}\frac{dl}{\frac{l^2}{2\mu}-E} = \frac{2\pi i\mu}{k(E)}\,,\qquad k(E)=+\sqrt{2\mu E}\,,
\end{equation}
plays a special role in the LSE framework. Considered as a function of the energy $E$, it has a branch cut along $E>0$ in the complex $E$-plane. For real positive $E$, the integration contour is to be approached from the upper complex plane, and $k(E)=+\sqrt{2\mu E}$ is the positive square root; otherwise, $k(E)\equiv k$ is fixed to be the square root with the positive imaginary part. If not stated otherwise, the variable $k$ will always stand for this square root. These requirements define the first or ``physical'' Riemann sheet. Considering the integral in Eq.~(\ref{eq:LSEloop}) as a function of complex $k$, and continuing analytically to the lower complex $k$-plane, amounts to the analytic continuation in the variable $E$ over the positive real $E$-axis, to the second Riemann sheet. These analytic properties of $I_{0}(E)$ will also show up in the solution of the LSE. The relation between $I_{0}(E)$ and the scalar relativistic two-point loop integral is discussed in App.~\ref{app:relloops}.


\section{LSE for symmetric double-delta potential}
\label{app:LSdoubledelta}

For a one-dimensional potential involving two delta functions,
\begin{equation}
V(x)=\frac{v}{2}\delta(x+d)+\frac{v}{2}\delta(x-d)\quad\Rightarrow\quad \langle q'|\hat{V}|q\rangle = \frac{v}{2\pi}\cos((q'-q)d)\,,
\end{equation}
 with $d>0$ and a real constant $v$, the ansatz
\begin{equation}\label{eq:Tsoldd}
\langle q\,'|\hat{\mathcal{T}}(E)|q\rangle \equiv \mathcal{T}(q',q;E) = \cos(q'd)\,\mathcal{T}_{0}(E)\cos(qd) + \sin(q'd)\,\mathcal{T}_{1}(E)\sin(qd)
\end{equation}
can be checked to solve the LSE for
\begin{equation}\label{eq:T01}
\mathcal{T}_{0}(E) = \frac{\frac{v}{2\pi}}{1+\frac{i\mu v}{2k}\left(1+e^{2ikd}\right)}\,,\quad \mathcal{T}_{1}(E) = \frac{\frac{v}{2\pi}}{1+\frac{i\mu v}{2k}\left(1-e^{2ikd}\right)}\,.
\end{equation}
Obviously, $\mathcal{T}_{0}(E)$ tends to the solution for a potential $\sim v\delta(x)$ for $d\rightarrow 0$, as it should.\\
We can define the analogue of on-shell s-and p-wave ``partial wave'' amplitudes by fixing $q=k$ and averaging over $q'=\pm k$,
\begin{equation}
\mathcal{T}_{\mathrm{s}}(E):=\mathcal{T}_{0}(E)\cos^2kd\,,\quad \mathcal{T}_{\mathrm{p}}(E):=\mathcal{T}_{1}(E)\sin^2kd\,.\label{eq:pwdd}
\end{equation}
They can be written in the ``K-Matrix'' form
\begin{eqnarray}
  \mathcal{T}_{\mathrm{s}}(E) &=& \left\lbrack K_{\mathrm{s}}^{-1}+\frac{2\pi i\mu}{k}\right\rbrack^{-1}\,,\quad K_{\mathrm{s}}= \frac{v\cos^2kd}{2\pi(1-\frac{\mu v}{k}\sin kd\cos kd)}\,,\label{eq:TsK}\\
  \mathcal{T}_{\mathrm{p}}(E) &=& \left\lbrack K_{\mathrm{p}}^{-1}+\frac{2\pi i\mu}{k}\right\rbrack^{-1}\,,\quad K_{\mathrm{p}}= \frac{v\sin^2kd}{2\pi(1+\frac{\mu v}{k}\sin kd\cos kd)}\,.\label{eq:TpK}
\end{eqnarray}
We also define transmission and reflection amplitudes $\tau$ and $\rho$ via
\begin{equation}
\tau(E) = 1-\frac{2\pi i\mu}{k}\mathcal{T}(+k,k;E)\,,\qquad \rho(E) = -\frac{2\pi i\mu}{k}\mathcal{T}(-k,k;E)\,,\label{eq:taurhodd}
\end{equation}
and note the relations $\tau+\rho-1=2\left(-\frac{2\pi i\mu}{k}\right)\mathcal{T}_{\mathrm{s}}\,$, $\,\tau-\rho-1=2\left(-\frac{2\pi i\mu}{k}\right)\mathcal{T}_{\mathrm{p}}\,$. Explicitly,
\begin{equation}\label{eq:taurhoddexplicit}
  \tau(E) = \left\lbrack 1+\frac{i\mu v}{k}-\frac{\mu^2v^2}{4k^2}\left(1-e^{4ikd}\right)\right\rbrack^{-1}\,,\quad i\rho(E)=\frac{\mu v}{k}\left(\cos 2kd + \frac{\mu v}{2k}\sin 2kd\right)\tau(E)\,.
\end{equation}
Unitarity can be expressed in the form $|\tau(E)\pm \rho(E)|=1$ for $E>0$. Setting $k=i\kappa$ with $\kappa>0$ in Eq.~(\ref{eq:T01}), one finds that there are bound-state poles in $\mathcal{T}_{0}$ for $v<0$ (and for $\mu vd<-1$ also in $\mathcal{T}_{1}$). Let us, therefore, first consider the case where $v>0$ (no bound states). The functions $\mathcal{T}_{0,1}-\frac{v}{2\pi}$ vanish as $E\rightarrow\infty$, and have a unitarity cut for $E>0$, but are analytic on the rest of the physical Riemann sheet ($\mathrm{Im}\,k>0$). Therefore we have the dispersion relations
\begin{equation}
\mathcal{T}_{i}(E) = \frac{v}{2\pi} + \frac{1}{\pi}\int_{0}^{\infty}dE'\,\frac{\mathrm{Im}\mathcal{T}_{i}(E')}{E'-E-i\epsilon}\,,\qquad i=0,1\,;\,\,v>0\,.
\end{equation}
Note that the physical real energy axis is to be approached from the upper complex $E$-plane. With the help of Eqs.~(\ref{eq:Tsoldd}), (\ref{eq:pwdd}), and (\ref{eq:taurhodd}),
we can rewrite these representations as
\begin{eqnarray}
  \mathcal{T}_{0}(E) &=& \frac{v}{2\pi} - \frac{1}{8\pi^2\mu^2}\int_{0}^{\infty}dk'\,\frac{k'^2\left|\frac{\tau(E')+\rho(E')-1}{\cos k'd}\right|^2}{E'-E-i\epsilon}\,,\\
  \mathcal{T}_{1}(E) &=& \frac{v}{2\pi} - \frac{1}{8\pi^2\mu^2}\int_{0}^{\infty}dk'\,\frac{k'^2\left|\frac{\tau(E')-\rho(E')-1}{\sin k'd}\right|^2}{E'-E-i\epsilon}\,,
\end{eqnarray}
where we have introduced $k':=+\sqrt{2\mu E'}$. Employing the results of App.~\ref{app:LSdoubledeltawf} for the wave functions pertaining to the double-delta potential, we can furthermore rewrite this in terms of continuum wave functions in momentum space as follows,
\begin{eqnarray*}
  \cos(q'd)\,\mathcal{T}_{0}(E)\cos(qd) &=& \cos(q'd)\frac{v}{2\pi}\cos(qd) - \int_{0}^{\infty}\frac{dk'}{2\pi}\,\frac{\left(\frac{q'^2}{2\mu}-E'\right)\tilde{\psi}^{+}_{E'}(q')\tilde{\psi}_{E'}^{+\ast}(q)\left(\frac{q^2}{2\mu}-E'\right)}{E'-E-i\epsilon}\,,\\
  \sin(q'd)\,\mathcal{T}_{1}(E)\sin(qd) &=& \sin(q'd)\frac{v}{2\pi}\sin(qd) - \int_{0}^{\infty}\frac{dk'}{2\pi}\,\frac{\left(\frac{q'^2}{2\mu}-E'\right)\tilde{\psi}^{-}_{E'}(q')\tilde{\psi}_{E'}^{-\ast}(q)\left(\frac{q^2}{2\mu}-E'\right)}{E'-E-i\epsilon}\,.
\end{eqnarray*}
In the more general case where there can be bound states, the integration contour on the physical sheet picks up pole terms, which are added to the continuum contribution. It is straightforward to compare the residues of the amplitudes $\mathcal{T}_{0,1}$ in Eq.~(\ref{eq:T01}) to the bound-state wave functions given in App.~\ref{app:LSdoubledeltawf}. For the complete amplitude of Eq.~(\ref{eq:Tsoldd}), we then find
\begin{eqnarray}
\mathcal{T}(q',q;E) &=& \frac{v}{2\pi}\cos((q'-q)d) + \,\theta(-v)\frac{\left(\frac{q'^2}{2\mu}-E_{B}^{+}\right)\tilde{\psi}^{+}_{B}(q')\tilde{\psi}_{B}^{+\ast}(q)\left(\frac{q^2}{2\mu}-E_{B}^{+}\right)}{2\pi(E-E_{B}^{+})} \nonumber \\ &+& \,\theta(-\mu vd-1)\frac{\left(\frac{q'^2}{2\mu}-E_{B}^{-}\right)\tilde{\psi}^{-}_{B}(q')\tilde{\psi}_{B}^{-\ast}(q)\left(\frac{q^2}{2\mu}-E_{B}^{-}\right)}{2\pi(E-E_{B}^{-})} \nonumber \\ &+& \int_{0}^{\infty}\frac{dk'}{2\pi}\,\frac{\left(\frac{q'^2}{2\mu}-E'\right)\tilde{\psi}^{+}_{E'}(q')\tilde{\psi}_{E'}^{+\ast}(q)\left(\frac{q^2}{2\mu}-E'\right)}{E+i\epsilon -E'} \nonumber \\ &+& \int_{0}^{\infty}\frac{dk'}{2\pi}\,\frac{\left(\frac{q'^2}{2\mu}-E'\right)\tilde{\psi}^{-}_{E'}(q')\tilde{\psi}_{E'}^{-\ast}(q)\left(\frac{q^2}{2\mu}-E'\right)}{E+i\epsilon -E'}\,,\label{eq:Trepwf}
\end{eqnarray}
compare also Eq.~(\ref{eq:LSGmat}). Here $E_{B}^{\pm}<0$ are the bound-state energies found from the poles $k_{B}^{\pm}=i\kappa^{\pm}_{B}$ in $\mathcal{T}_{0}$ and $\mathcal{T}_{1}$, where $\kappa^{\pm}_{B}=+\sqrt{-2\mu E_{B}^{\pm}}\,$.


\section{Form factors}
\label{app:FF}

\begin{figure}[h]
\centering
\subfigure[\,$D_{R}^{(+)}(x)$]{\includegraphics[width=0.50\textwidth]{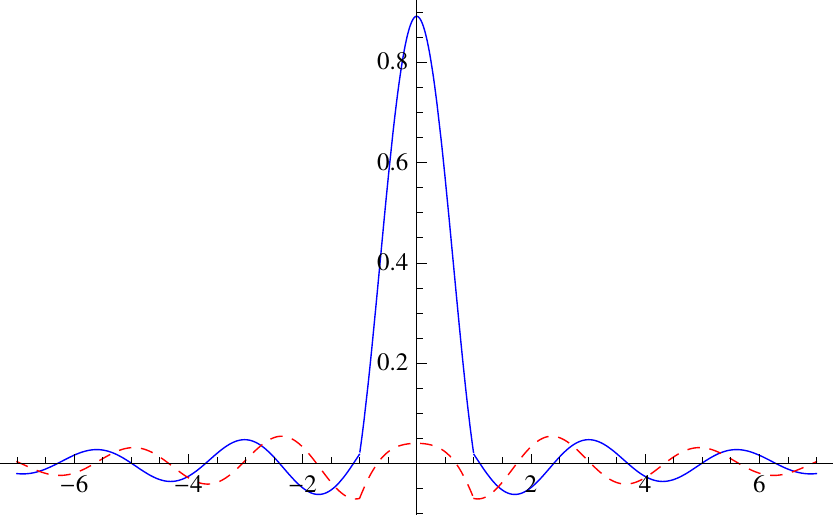}}\\
\subfigure[\,$D_{R}^{(-)}(x)$]{\includegraphics[width=0.50\textwidth]{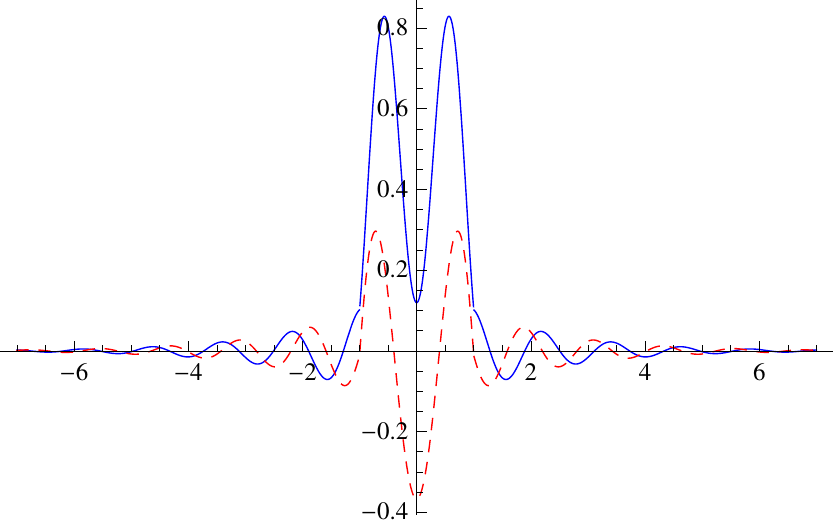}}
\caption{Real (blue) and imaginary (red, dashed) part of the density functions $D_{R}^{(\pm)}(x)$\,, for $v=3$, $\mu=d=1$. The resonance poles are located at $k_{R}^{+}\approx 1.213-0.105\,i$ and $k_{R}^{-}\approx 2.572-0.317\,i$, respectively.}
\label{fig:DRpm}
\end{figure}

\begin{figure}[h]
\centering
\subfigure[\,$D_{R}^{(+)}(x)$]{\includegraphics[width=0.45\textwidth]{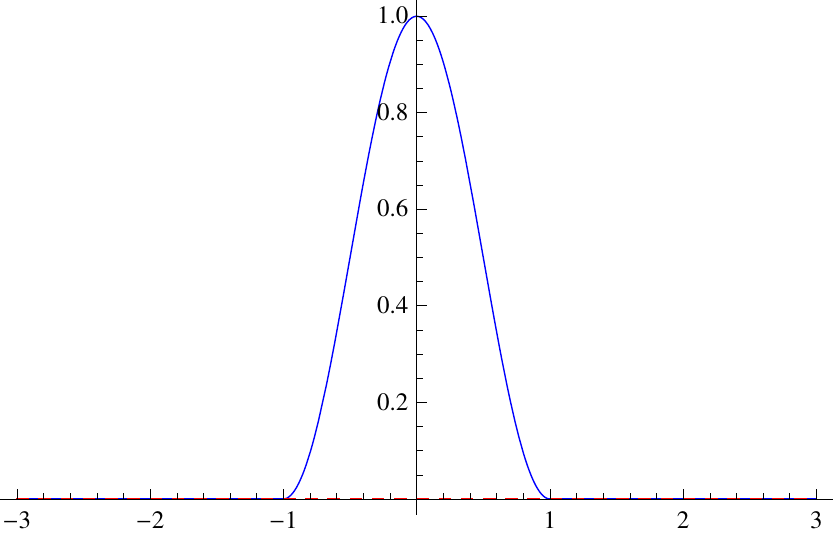}}
\subfigure[\,$D_{R}^{(-)}(x)$]{\includegraphics[width=0.45\textwidth]{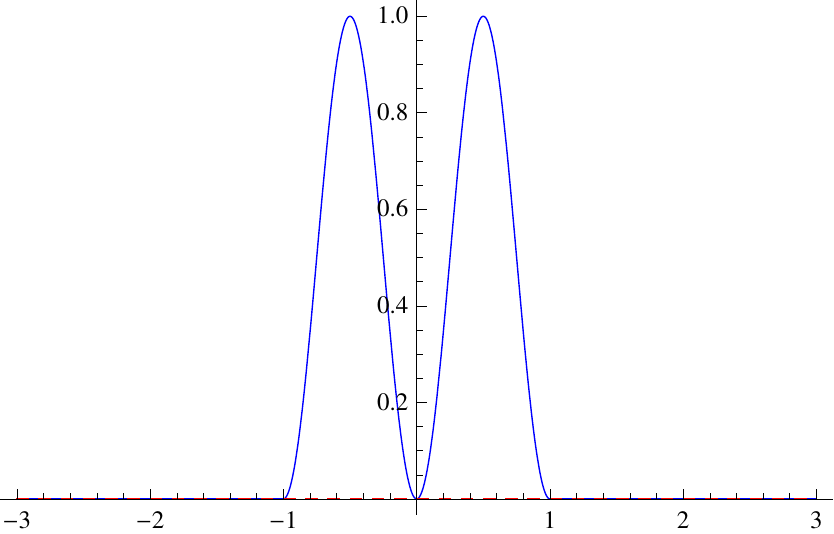}}
\caption{The density functions $D_{R}^{(\pm)}(x)$\, of the lowest-lying resonances for $\mu=d=1$, in the limit of infinite repulsive coupling strenth $v$. The poles are located at $k_{R}^{+}=\frac{\pi}{2d}$ and $k_{R}^{-}=\frac{\pi}{d}$, respectively.}
\label{fig:DRpminf}
\end{figure}

Given the bound state wave functions of the previous section, we can define the form factor of a bound state $B$ as
\begin{eqnarray}
  F_{B}(Q^2) &:=& \int_{-\infty}^{\infty}dx\,\psi^{\ast}_{B}(x)e^{iQx}\psi_{B}(x) = 1-\frac{1}{2}Q^2\langle\hat{x}^2\rangle_{B} + \mathcal{O}(Q^4)\,,\label{eq:defF}\\
  \langle\hat{x}^2\rangle_{B} &:=& \int_{-\infty}^{\infty}dx\,\psi^{\ast}_{B}(x)x^2\psi_{B}(x)\,.\label{eq:defxsqr}
\end{eqnarray}
In terms of the momentum-space wave functions, the form factor can be rewritten as
\begin{eqnarray}
  F_{B}(Q^2) &=& \int_{-\infty}^{\infty}\frac{dp}{2\pi}\,\tilde{\psi}^{\ast}_{B}(p+Q)\tilde{\psi}_{B}(p) \nonumber \\
  &=& \int_{-\infty}^{\infty}\frac{dp}{2\pi}\,\frac{\tilde{\psi}^{\ast}_{B}(p+Q)\left(\frac{(p+Q)^2}{2\mu}-E_{B}\right)\left(\frac{p^2}{2\mu}-E_{B}\right)\tilde{\psi}_{B}(p)}{\left[\frac{(p+Q)^2}{2\mu}-E_{B}\right]\left[\frac{p^2}{2\mu}-E_{B}\right]}\,.
\end{eqnarray}
Making use of the representation of the scattering amplitude in terms of the wave functions, Eq.~(\ref{eq:Trepwf}), we can rewrite this further in terms of the residuum of the scattering amplitude at the bound-state pole:
\begin{equation}
  F_{B}(Q^2) = \frac{1}{\mathrm{Res}\,\mathcal{T}(q',q;E\rightarrow E_{B})}\int_{-\infty}^{\infty}dp\,\frac{\left(\mathrm{Res}\,\mathcal{T}(q',p+Q;E\rightarrow E_{B})\right)\left(\mathrm{Res}\,\mathcal{T}(p,q;E\rightarrow E_{B})\right)}{\left[\frac{(p+Q)^2}{2\mu}-E_{B}\right]\left[\frac{p^2}{2\mu}-E_{B}\right]}\,.\label{eq:FFgeneral}
\end{equation}
The integral
\begin{equation}\label{eq:F0}
\mathcal{F}^{0}(Q^2;E):=\int_{-\infty}^{\infty}\frac{dp}{\left[\frac{(p+Q)^2}{2\mu}-E\right]\left[\frac{p^2}{2\mu}-E\right]} = \frac{8\pi i\mu^2}{k(Q^2-4k^2)}\,
\end{equation}
can be related to the non-relativistic limit of the relativistic loop integral with three propagators in $d=1+1$ dimensions, compare App.~\ref{app:relloops}.
Moreover, we shall make use of
\begin{eqnarray}
  \mathcal{F}^{+}(Q^2;E) &:=& \int_{-\infty}^{\infty}dp\,\frac{\cos((p+Q)d)\cos pd}{\left[\frac{(p+Q)^2}{2\mu}-E\right]\left[\frac{p^2}{2\mu}-E\right]} \nonumber \\ &=& \frac{8\pi i\mu^2}{k(Q^2-4k^2)}\left(e^{-ikd}\cos kd\cos Qd - ikd\left(\frac{\sin Qd}{Qd}\right)\right)e^{2ikd}\,,\\
  \mathcal{F}^{-}(Q^2;E) &:=& \int_{-\infty}^{\infty}dp\,\frac{\sin((p+Q)d)\sin pd}{\left[\frac{(p+Q)^2}{2\mu}-E\right]\left[\frac{p^2}{2\mu}-E\right]} \nonumber \\ &=& \frac{8\pi i\mu^2}{k(Q^2-4k^2)}\left(-ie^{-ikd}\sin kd\cos Qd + ikd\left(\frac{\sin Qd}{Qd}\right)\right)e^{2ikd}\,.
\end{eqnarray}
Note again that real positive values of $E$ are to be approached from the upper complex plane here, and that $\mathrm{Im}\,k\geq 0$.
For example, we find for the s-wave bound state (compare Eq.~(\ref{eq:SchrBplus})):
\begin{eqnarray}
  F_{B}^{(+)}(Q^2) &=& \frac{v^2}{2\pi}(\mathcal{N}_{B}^{+})^2e^{-2\kappa_{B}d}\mathcal{F}^{+}(Q^2;E_{B}) \\
  &=& \frac{8\kappa_{B}^2e^{-2\kappa_{B}d}\left(e^{\kappa_{B}d}\cosh\kappa_{B}d\cos Qd + \kappa_{B}d\left(\frac{\sin Qd}{Qd}\right)\right)}{(4\kappa_{B}^2+Q^2)(1+(1+2\kappa_{B}d)e^{-2\kappa_{B}d})}\,,\nonumber\\
  \langle\hat{x}^2\rangle_{B}^{(+)} &=& d^2+\frac{1}{2\kappa_{B}^2} -\frac{4\kappa_{B}d^3e^{-2\kappa_{B}d}}{3(1+(1+2\kappa_{B}d)e^{-2\kappa_{B}d})}\,,\label{eq:xpsqr}
\end{eqnarray}
and for the p-wave bound state
\begin{eqnarray}
  F_{B}^{(-)}(Q^2) &=& \frac{v^2}{2\pi}(\mathcal{N}_{B}^{-})^2e^{-2\kappa_{B}d}\mathcal{F}^{-}(Q^2;E_{B}) \\
  &=& \frac{8\kappa_{B}^2e^{-2\kappa_{B}d}\left(e^{\kappa_{B}d}\sinh\kappa_{B}d\cos Qd - \kappa_{B}d\left(\frac{\sin Qd}{Qd}\right)\right)}{(4\kappa_{B}^2+Q^2)(1-(1+2\kappa_{B}d)e^{-2\kappa_{B}d})}\,,\nonumber\\
  \langle\hat{x}^2\rangle_{B}^{(-)} &=& d^2+\frac{1}{2\kappa_{B}^2} + \frac{4\kappa_{B}d^3e^{-2\kappa_{B}d}}{3(1-(1+2\kappa_{B}d)e^{-2\kappa_{B}d})}\,,\label{eq:xmsqr}
\end{eqnarray}
where $\kappa_{B}=\sqrt{-2\mu E_{B}}$ is given by the pole position of the respective bound state ($E_{B}<0$). One can verify that the inverse Fourier transforms of these form factors yield the absolute squares of the corresponding bound-state wave functions of Eqs.~(\ref{eq:psiBplus}),\,(\ref{eq:psiBminus}),
\begin{equation}\label{eq:psiBcheck}
\int_{-\infty}^{\infty}\frac{dQ}{2\pi}\,e^{-iQx}F_{B}^{(\pm)}(Q^2) = \psi^{\pm\ast}_{B}(x)\psi_{B}^{\pm}(x)\,.
\end{equation}
At first sight, it seems straightforward to extend this procedure to the case of resonances. Eq.~(\ref{eq:FFgeneral}) suggests that the ``resonance form factor'' could be extracted from the extrapolation to the double-pole term of an amplitude describing a scattering process in the presence of an external field, which transfers a momentum $\sim Q$ to the system (see also \cite{Gegelia:2009py,Bernard:2012bi}). Indeed, the functions $\mathcal{F}^{\pm}(Q^2;E)$ and the scattering amplitude $\mathcal{T}$ are given as explicit functions of $k(E)$, and their analytic continuation to the second Riemann sheet can directly be inferred by considering them as functions of complex $k$, and continuing these functions analytically to the lower $k$-plane. Let $k_{R}^{\pm}$ denote resonance pole positions in this lower half-plane ($\mathrm{Re}\,k_{R}^{\pm}>0$, $\mathrm{Im}\,k_{R}^{\pm}<0\,$). Continuing the relation of Eq.~(\ref{eq:FFgeneral}) to the poles in the lower $k$-plane, which we might consider a natural definition of ``resonance form factors'' $F_{R}$, leaves us with
\begin{eqnarray}
  F_{R}^{(+)}(Q^2) &=& -\frac{8k_{R}^2e^{2ik_{R}d}\left(e^{-ik_{R}d}\cos k_{R}d\cos Qd - ik_{R}d\left(\frac{\sin Qd}{Qd}\right)\right)}{(Q^2-4k_{R}^2)(1+(1-2ik_{R}d)e^{2ik_{R}d})}\,,\label{eq:FRplus}\\
  F_{R}^{(-)}(Q^2) &=& -\frac{8k_{R}^2e^{2ik_{R}d}\left(-ie^{-ik_{R}d}\sin k_{R}d\cos Qd + ik_{R}d\left(\frac{\sin Qd}{Qd}\right)\right)}{(Q^2-4k_{R}^2)(1-(1-2ik_{R}d)e^{2ik_{R}d})}\,,\label{eq:FRminus}
\end{eqnarray}
with $k_{R}$ set to the respective fixed resonance position $k_{R}^{\pm}$. Defining ``resonance densities'' $D_{R}$ via the inverse Fourier transforms of these form factors,
\begin{equation}
D_{R}^{(\pm)}(x) := \int_{-\infty}^{\infty}\frac{dQ}{2\pi}\,e^{-iQx}F_{R}^{(\pm)}(Q^2)\,,
\end{equation}
leads us to the following functions, examples of which are plotted in Figs.~\ref{fig:DRpm} and \ref{fig:DRpminf}:\\
%
%
\begin{eqnarray}
  D_{R}^{(+)}(x) &=& (\mathcal{N}^{+}_{R})^2\biggl(\theta(-d-x)e^{2ik_{R}x} + \theta(d-x)\theta(x+d)\left(\frac{(2+e^{-2ik_{R}d})\cos2k_{R}x -e^{2ik_{R}d}}{2\cos^2k_{R}d}\right) \nonumber \\ & & \hspace{4.4cm} + \,\theta(x-d)e^{-2ik_{R}x}\biggr)\,,\label{eq:DRplus}\\
  D_{R}^{(-)}(x) &=& (\mathcal{N}^{-}_{R})^2\biggl(\theta(-d-x)e^{2ik_{R}x} + \theta(d-x)\theta(x+d)\left(\frac{(2-e^{-2ik_{R}d})\cos2k_{R}x -e^{2ik_{R}d}}{2\sin^2k_{R}d}\right) \nonumber \\ & & \hspace{4.4cm} + \,\theta(x-d)e^{-2ik_{R}x}\biggr)\,,\label{eq:DRminus}\\
  \mathcal{N}^{+}_{R} &=& \frac{\sqrt{2ik_{R}}\cos k_{R}d}{\sqrt{1+(1-2ik_{R}d)e^{2ik_{R}d}}}\,,\qquad \mathcal{N}^{-}_{R} = \frac{\sqrt{-2ik_{R}}\sin k_{R}d}{\sqrt{1-(1-2ik_{R}d)e^{2ik_{R}d}}}\,.
\end{eqnarray}
One can directly verify that these densities are normalized to one, and yield
\begin{eqnarray}
  \int_{-\infty}^{\infty}dx\,x^2D_{R}^{(+)}(x) &=& d^2-\frac{1}{2k_{R}^2} + \frac{4ik_{R}d^3e^{2ik_{R}d}}{3(1+(1-2ik_{R}d)e^{2ik_{R}d})}\,,\\
  \int_{-\infty}^{\infty}dx\,x^2D_{R}^{(-)}(x) &=& d^2-\frac{1}{2k_{R}^2} - \frac{4ik_{R}d^3e^{2ik_{R}d}}{3(1-(1-2ik_{R}d)e^{2ik_{R}d})}\,,
\end{eqnarray}
which are indeed the analytic continuations of the results (\ref{eq:xpsqr}) and (\ref{eq:xmsqr}). Note, however, that the above densities are {\em not\,} the analytic continuations of the (absolute) squares of the bound state wave functions of Eqs.~(\ref{eq:psiBplus}), (\ref{eq:psiBminus})\,. The latter are not normalizable (exponentially increasing as $|x|\rightarrow\infty$, since $\mathrm{Im}\,k_{R}<0$) - in particular, they cannot yield well-defined finite mean-square radii in the sense of Eq.~(\ref{eq:defxsqr}). The term ``resonance wave function'' is reserved for those non-normalizable analytic continuations of the bound-state wave functions in the literature \cite{Gamow:1928zz,Zeldovich:1961a,Berggren:1968zz,Garcia-Calderon:1976omn}. Though it is tempting to interprete the densities $D_{R}^{(\pm)}(x)$ as squares of a wave function of a particle ``trapped'' for some time between the potential barriers, this is not possible without some drawbacks: for example, while $\psi^{-\ast}_{B}(x)\psi^{-}_{B}(x)$ vanishes at $x=0$, reflecting the odd parity of the bound-state wave function $\psi^{-}_{B}(x)$, we have in general $D_{R}^{(-)}(0)\not=0$. The reason for this somewhat unexpected behavior is the pole term $\sim(Q^2-4k^2)^{-1}$ stemming from the ``anomalous threshold'' part of the triangle integral (\ref{eq:IDelta}): the analytic continuation of integrals like (\ref{eq:psiBcheck}) has to cross the real $k$-axis, on which the poles in the $Q$-integration are located.\\

\section{Description with an energy-dependent potential}
\label{sec:edepv}
\begin{figure}[b]
\centering
\includegraphics[width=0.65\textwidth]{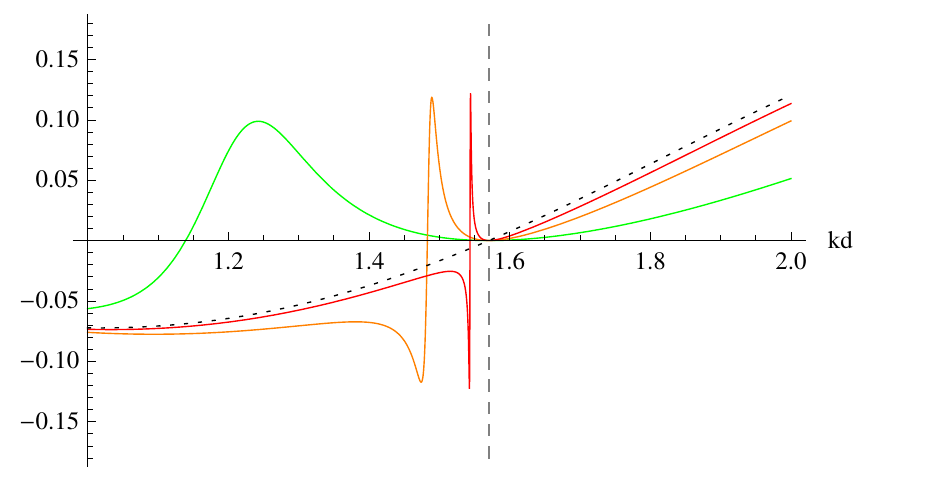}
\caption{Real part of $\mathcal{T}_{\mathrm{s}}$ for $v=3$ (green), $v=17$ (orange) and $v=57$ (red), $\mu=d=1$. The gray dashed line marks $kd=\frac{\pi}{2}$, the black dotted line results in the limit $v\rightarrow\infty$.}
\label{fig:Tslimit}
\end{figure}

The ``partial-wave'' amplitude $\mathcal{T}_{s}(E)$ of Eq.~(\ref{eq:TsK}) can also result as the solution of a LSE with an energy-dependent potential of the form
\begin{equation}\label{eq:vsE}
V(x)=v_{\mathrm{s}}(E)\delta(x)\,,\quad v_{\mathrm{s}}(E) = 2\pi K_{\mathrm{s}}(E)\,.
\end{equation}
By construction, all pole positions and residues in the complex energy plane resulting from this LSE-kernel are the same as those of $\mathcal{T}_{s}(E)$.
We still define the bound-state wave functions $\tilde{\Phi}_{B}$ (up to a phase) by requiring that the bound-state pole in the LSE-solution $\mathcal{T}(q',q;E)$ is of the form
\begin{displaymath}
\frac{\left(\frac{q'^2}{2\mu}-E_{B}\right)\tilde{\Phi}_{B}(q')\tilde{\Phi}_{B}^{\ast}(q)\left(\frac{q^2}{2\mu}-E_{B}\right)}{2\pi(E-E_{B})}\,,
\end{displaymath}
compare Eq.~(\ref{eq:Trepwf}). Of course, the form of the bound-state and continuum solutions, denoted as $\Phi_{B}(x)\,,\,\Phi_{E}(x)$, respectively, can immediately be obtained from the functions $\psi_{B}^{+}(x)\,,\,\psi_{E}^{+}(x)$ of App.~\ref{app:LSdoubledeltawf}, taking the limit $d\rightarrow 0$ there, and replacing $v\rightarrow v_{\mathrm{s}}(E)$. However, as is well-known \cite{Lepage:1977gd,Sazdjian:1986qn,Mares:2004}, the scalar product of the wave functions has to be modified for energy-dependent potentials:
\begin{equation}\label{eq:newsp}
\langle \Phi_{E_{1}}\,,\,\Phi_{E_{2}}\rangle_{V'}:= \int_{-\infty}^{\infty}dy\int_{-\infty}^{\infty}dx\,\Phi^{\ast}_{E_{1}}(y)\left\lbrack\delta(x-y) - \left\langle y\left|\frac{\hat{V}(E_{1})-\hat{V}(E_{2})}{E_{1}-E_{2}}\right|x\right\rangle\right\rbrack\Phi_{E_{2}}(x)\,.
\end{equation}
Then, eigenfunctions of different energies are orthogonal in the sense that $\langle \Phi_{E_{1}}\,,\,\Phi_{E_{2}}\rangle_{V'}=0$ for $E_{1}\not=E_{2}$, while the normalization conditions ($E_{2}\rightarrow E_{1}$) contain an integral involving $\frac{\partial\hat{V}}{\partial{E}}\bigr|_{E_{1}}\,$. These non-standard normalization conditions are automatically fulfilled if one defines the wave functions via a representation of a solution to a LSE as in Eq.~(\ref{eq:Trepwf}), compare Eq.~(\ref{eq:LSEBnorm}). \\ For the bound-state wave functions in our example, we obtain
\begin{eqnarray}
  \Phi_{B}(x) &=& \mathcal{N}_{B}^{+}e^{-\kappa_{B}|x|}\,,\quad \int_{-\infty}^{\infty}dx\,\Phi^{\ast}_{B}(x)\Phi_{B}(x) = \frac{(\mathcal{N}_{B}^{+})^2}{\kappa_{B}}>1\quad\mathrm{for}\quad d>0\,,\\
  1 &=& \left(\int_{-\infty}^{\infty}dx\,\Phi^{\ast}_{B}(x)\Phi_{B}(x)\right) - (\mathcal{N}_{B}^{+})^2\frac{\partial v_{\mathrm{s}}}{\partial E}\biggr|_{E_{B}} = \langle \Phi_{B}\,,\,\Phi_{B}\rangle_{V'}\,,\quad E_{B}=-\frac{\kappa_{B}^2}{2\mu}\,.
\end{eqnarray}
See App.~\ref{app:LSdoubledeltawf} for the normalization factor $\mathcal{N}_{B}^{+}\,$. Expectation values also involve the modified scalar product. For example, we generalize the definition of the bound-state form factor to
\begin{eqnarray}
  F_{B}^{(\mathrm{s})}(Q^2) &:=& \langle \Phi_{B}\,,\,e^{iQ\hat{x}}\Phi_{B}\rangle_{V'}  \label{def:FFBEdep}\\
  &=& \int_{-\infty}^{\infty}dx\,\Phi^{\ast}_{B}(x)e^{iQx}\Phi_{B}(x) - \int_{-\infty}^{\infty}dy\int_{-\infty}^{\infty}dx\,\Phi^{\ast}_{B}(y)\left\langle y\left|\frac{\partial\hat{V}}{\partial E}(E_{B})\right|x\right\rangle e^{iQx}\Phi_{B}(x)\,.\nonumber
\end{eqnarray}
Applied to our specific example, this definition results in
\begin{equation}\label{eq:FFBEdep}
F_{B}^{(\mathrm{s})}(Q^2) = (\mathcal{N}_{B}^{+})^2\left(\frac{4\kappa_{B}}{4\kappa_{B}^2+Q^2}-\frac{\partial v_{\mathrm{s}}}{\partial E}\biggr|_{E_{B}}\right) = \frac{\frac{4\kappa_{B}}{4\kappa_{B}^2+Q^2}-\frac{\partial v_{\mathrm{s}}}{\partial E}\bigr|_{E_{B}}}{\frac{1}{\kappa_{B}}-\frac{\partial v_{\mathrm{s}}}{\partial E}\bigr|_{E_{B}}}\,.
\end{equation}
In particular, $F_{B}^{(\mathrm{s})}(0)=1$, and all the moments (expectation values) $\langle \hat{x}^{2n}\rangle_{B}^{(\mathrm{s})}$ are positive. E.g., $\langle \hat{x}^{2}\rangle_{B}^{(\mathrm{s})}=(\mathcal{N}_{B}^{+})^2/(2\kappa_{B}^3) > \langle \hat{x}^{2}\rangle_{B}^{(+)}$ for $d>0$, compare (\ref{eq:xpsqr})\,.
Note that the ratio $\langle \hat{x}^{2}\rangle_{B}^{(\mathrm{s})}/\langle \hat{x}^{2}\rangle_{B}^{(+)}$ grows {\em exponentially\,} with $\kappa_{B}d$. So even if we have a perfect description for the (on-shell, i.e. $|q'|=|q|=k$) ``partial-wave'' amplitude $\mathcal{T}_{s}(E)$, the wave functions and form factors derived from this description can be totally wrong. The form factors contain additional information, which can not be inferred from on-shell partial-wave scattering amplitudes.\\
On the other hand, if the bound state is located very close to the threshold (small $\kappa_{B}\ll\frac{1}{d}$), both results for the expectation value approach the ``universal'' result $\langle \hat{x}^{2}\rangle\rightarrow (2\kappa_{B}^2)^{-1}\,$.\\
\newpage
The potential $v_{\mathrm{s}}(E)$ can have poles for positive energies $E_{0}$. Very close to such a pole, we can approximate
\begin{equation}\label{eq:vsapprox}
v_{\mathrm{s}}(E) \approx v_{0} + \frac{g^2}{E-E_{0}}\,,
\end{equation}
where the parameters $v_{0},\,g^2$ are given by
\begin{eqnarray}
  v_{0} &=& 2k_{0}\cos^2k_{0}d\,\frac{\sin 2k_{0}d(1-2(k_{0}d)^2) -2k_{0}d\cos 2k_{0}d}{\mu(\sin 2k_{0}d -2k_{0}d\cos 2k_{0}d)^2}\,,\\
  g^2 &=& \frac{2k_{0}^3\cos^2k_{0}d}{\mu^2(\sin 2k_{0}d -2k_{0}d\cos 2k_{0}d)}\,,\qquad k_{0}=\sqrt{2\mu E_{0}}\,.
\end{eqnarray}
Note, though, that $g^2$ is not necessarily positive here (but for sufficiently small $k_{0}d$, it is). {\em Given\,} a potential of the form of the r.h.s. of Eq.~(\ref{eq:vsapprox}), the second term on the r.h.s. of Eqs.~(\ref{def:FFBEdep}), (\ref{eq:FFBEdep}), $\sim g^2/(E-E_{0})^2$, offers a nice interpretation: it describes a coupling of the probe to a ``bare'' point particle of mass $4\mu+E_{0}$, exchanged between vertices with amplitudes given by $\sim g\mathcal{N}_{B}^{+}$, while the first term can again be interpreted in terms of the triangle integral describing the coupling of the probe to a pair of particles, as in (\ref{eq:FFgeneral}). Generally, the second term in (\ref{def:FFBEdep}) could be interpreted as the probe coupling to ``anything else than the particle pair''. \\
\quad \\
Let us see what happens in our model (\ref{eq:Tsoldd}) when the first resonance pole (at $k_{R}$) in $\mathcal{T}_{0}$ approaches such a $k_{0}$ (this happens when $v$ assumes large positive values, so that $k_{0}d\rightarrow\frac{\pi}{2} \leftarrow k_{R}d$). One observes that the resonance peak in $\mathcal{T}_{0}$ becomes higher and more and more narrow, until it decouples completely from the on-shell scattering process when the width goes to zero and $\cos k_{R}d\rightarrow 0$. See Fig.~\ref{fig:Tslimit} for corresponding plots of $\mathrm{Re\,}\mathcal{T}_{\mathrm{s}}$ over $kd$. \\
The ``resonance form factor'' of Eq.~(\ref{eq:FRplus}) in this limiting case tends to
\begin{equation}\label{eq:FRlim}
  F_{R}^{(+)}(Q^2) \rightarrow \frac{4k_{R}^2}{4k_{R}^2-Q^2}\left(\frac{\sin Qd}{Qd}\right)\,,\quad k_{R}\rightarrow \frac{\pi}{2d}\,,
\end{equation}
and we find for the pertinent density function:
\begin{equation}\label{eq:denslim}
  D_{R}^{(+)}(x) \rightarrow \theta(d-x)\theta(x+d)\frac{1}{d}\cos^2\left(\frac{\pi x}{2d}\right)\,, \qquad \int_{-d}^{d}dx\,D_{R}^{(+)}(x)\rightarrow 1\,,
\end{equation}
which seems to describe a particle which cannot escape to the outside region $|x|>d$ (or, in the overall picture, a particle pair which cannot have a distance greater than $d$ from one another).\\
See Fig.~\ref{fig:DRpminf}(a) for a corresponding plot of $D_{R}^{(+)}(x)$.\\
\quad \\
\newpage
Returning to the decription in terms of an energy-dependent potential, the analytic continuation of Eq.~(\ref{eq:FFBEdep}) to the lower complex $k$-plane is given by
\begin{eqnarray}
F_{R}^{(\mathrm{s})}(Q^2) &=&  \frac{\frac{4k_{R}^2}{4k_{R}^2-Q^2}+ik_{R}\frac{\partial v_{\mathrm{s}}}{\partial E}\bigr|_{E_{R}}}{1+ik_{R}\frac{\partial v_{\mathrm{s}}}{\partial E}\bigr|_{E_{R}}} \nonumber \\ &\approx& \frac{k_{R}g^2}{k_{R}g^2+i(E_{R}-E_{0})^2} + \frac{4i(E_{R}-E_{0})^2k_{R}^2}{(4k_{R}^2-Q^2)(k_{R}g^2+i(E_{R}-E_{0})^2)}\,,\label{eq:FFREdep}
\end{eqnarray}
where in the second line we employed the approximate description of Eq.~(\ref{eq:vsapprox}), with $E_{R}=\frac{k_{R}^2}{2\mu}$. Neglecting terms of $\mathcal{O}((E_{R}-E_{0})^2)$, the first term in the above decomposition is equal to one (in our model, it behaves as $\sim 1-\mathcal{O}(\frac{1}{v^2})$ in the limit $v\rightarrow\infty$), while the second one is zero. On the other hand, for $g\rightarrow 0$ with $E_{R}\not=E_{0}$, the first term vanishes, while the second term gives the typical form factor for an energy- and momentum-independent potential (compare Eq.~(\ref{eq:FRplus}) in the limit $d\rightarrow 0$), showing an enhancement near $Q^2=8\mu\mathrm{Re}(E_{R})$. In the first case, the form factor approaches a $Q^2$-independent constant (unity), yielding exactly a delta-function as the pertinent density, obviously describing a structureless point particle. \\
In studies of resonances employing effective energy-dependent potentials parameterized as in Eq.~(\ref{eq:vsapprox}), the analysis of the second line of Eq.~(\ref{eq:FFREdep}) (for $Q^2=0$) is known as the decomposition into ``elementariness'' (first term) and ``compositeness'' (second term) \cite{Zeldovich:1961b,Salam:1962ap,Weinberg:1962hj,Weinberg:1965zz,Baru:2003qq,Pelaez:2003dy,Baru:2010ww,Aceti:2012dd,Hyodo:2013nka,Sekihara:2014kya,Guo:2017jvc}. In general, the two complex numbers sum up to one for $Q^2=0$, where the first term gives the coefficient of the $\delta(x)$-part in the spatial distribution associated with the resonance. For the case of bound states, it is sometimes said that the ``elementariness'' gives the probability of ``{\em finding\,} a (bare) elementary state'' in the bound state\footnote{Zel'dovich \cite{Zeldovich:1961b} compares this kind of analysis with finding the fraction of iron in a mixture of iron and wood fillings by placing the mixture in a magnetic field.}. The concrete measurement implicit in this phrase can evidently just be taken as the measurement of form factors of the full states, yielding information about the spatial long- and short-distance structure. \\
The form factor of the model studied here, however, shows some $Q$-dependence even in the limiting case of Eq.~(\ref{eq:FRlim}). Only for $d\rightarrow 0$ does the corresponding density of Eq.~(\ref{eq:denslim}) approach a delta function. Thus, it depends on the properties of the ``real'' potential under study (or, more generally, on the real dynamics of the scattering process) how accurate and conclusive the description in terms of effective energy-dependent potentials can be.


\section{Conclusion}
\label{sec:Conclusio}

In this work, we have studied the properties of the bound states and resonances in a very simple quantum-mechanical model. The two ``partial-wave'' amplitudes (amplitudes of even or odd parity) can be written in a typical $K$-matrix form, with a $K$-function that shows a non-trivial energy dependence, including the possibility of poles (see Eqs.~(\ref{eq:TsK},\ref{eq:TpK})). The form factors are derived from the off-shell solution of the LSE, and their Fourier transforms lead to the expected density $\sim\psi^{\ast}_{B}(x)\psi_{B}(x)$ for the bound states $B$, while they lead to complex density functions $D_{R}(x)$ for the case of resonances $R$. We find that the latter can not be plainly interpreted as squares of pertinent wave functions - in some special cases, however, they still seem to give a nice intuitive picture of ``what is going on'' in the resonance region (see e.g. Fig.~\ref{fig:DRpminf}).
What appears as a particle ``trapped'' between two repulsive delta-functions, is described as a point-like (or almost point-like) particle when employing the description in terms of an effective energy-dependent potential (which has the the same pole positions and residues as the parity-even part of the complete (on-shell) amplitude  $\mathcal{T}_{\mathrm{s}}(E)$). In the case of increasing repulsive coupling strength, the ``elementariness'' (as inferred from the normalization of the wave functions, or from the form factor at $Q^2=0$) tends to one, while its counterpart, the ``compositeness'', tends to zero. If the spatial extent $d$ tends to zero, the two descriptions agree, while the full amplitude allows for a somewhat more detailed description of the structure of the states for finite $d$. From the simple one-dimensional example studied in this work, we conclude that it might be worthwhile to study ``resonance form factors'', e.g. from scattering processes probed by an external field that resolves the detailed structure of the scattering process in the resonance region, or maybe from conveniently chosen three-point functions evaluated in lattice QCD. Of course, one will have to rely on some parameterization that allows for an analytic continuation of the corresponding amplitude to the resonance poles. It will be important for such studies to minimize the model dependence of such parameterizations as far as possible. 


\begin{appendix}

\section{Wave functions for the double-delta potential}
\label{app:LSdoubledeltawf}
\def\theequation{\Alph{section}.\arabic{equation}}
\setcounter{equation}{0}
Let us first treat the continuum solutions ($E>0$, $k=+\sqrt{2\mu E}>0$). 
Solutions of type $L$ (``incoming particle from the left'') can be written as
\begin{eqnarray}
  \psi_{E}^{L}(x) &=& \frac{1}{\sqrt{2\pi}}\biggl(\theta(-d-x)\left(e^{ikx}+\rho(E)e^{-ikx}\right) + \theta(d-x)\theta(x+d)\left(a(E)e^{ikx}+b(E)e^{-ikx}\right) \nonumber \\
  & & \hspace{6.45cm} +\,  \theta(x-d)\tau(E)e^{ikx}\biggr)\,,\label{eq:psiLdd} \\
a(E) &=& 1+\frac{(\tau(E)-1)e^{2ikd}-\rho(E)}{2i\sin 2kd}\,,\quad b(E) = \frac{\rho(E)e^{2ikd}-(\tau(E)-1)}{2i\sin 2kd}\,.\quad \nonumber
\end{eqnarray}
Solutions of type $R$ (``incoming particle from the right''):
\begin{eqnarray}
\psi_{E}^{R}(x) &=& \frac{1}{\sqrt{2\pi}}\biggl(\theta(-d-x)\tau(E)e^{-ikx} + \theta(d-x)\theta(x+d)\left(a(E)e^{-ikx}+b(E)e^{ikx}\right) \nonumber \\
  & & \hspace{6.45cm} +\,  \theta(x-d)\left(e^{-ikx}+\rho(E)e^{ikx}\right)\biggr)\,.\label{eq:psiRdd}
\end{eqnarray}
The transmission and reflection amplitudes $\tau$ and $\rho$ are given explicitly in Eq.~(\ref{eq:taurhoddexplicit}).
One can verify that $a$ and $b$ stay finite when $2kd$ is a multiple of $\pi$ (the even multiples are zeroes of $\mathcal{T}_{\mathrm{p}}$, while the odd multiples are zeroes of $\mathcal{T}_{\mathrm{s}}\,$, see Eq.~(\ref{eq:pwdd})). - We can construct solutions of definite (even/odd) parity as follows,
\begin{equation}\label{eq:psipmdd}
\psi_{E}^{+}(x) = \frac{1}{\sqrt{2}}(\psi_{E}^{L}(x)+\psi_{E}^{R}(x))\,,\qquad \psi_{E}^{-}(x) = \frac{1}{\sqrt{2}}(\psi_{E}^{L}(x)-\psi_{E}^{R}(x))\,,
\end{equation}
which have the Fourier transforms
\begin{eqnarray*}
  \tilde{\psi}_{E}^{+}(p) &:=& \int_{-\infty}^{\infty}dx\,e^{-ipx}\psi_{E}^{+}(x) = \frac{1}{\sqrt{4\pi}}\biggl(2\pi\delta(p-k)+2\pi\delta(p+k)  \\
  &+& ie^{ikd}\left(\frac{\tau(E)+\rho(E)-1}{\cos kd}\right)\left(\frac{e^{ipd}\cos kd - i\sin(p+k)d}{p+k+i\epsilon} - \frac{e^{-ipd}\cos kd + i\sin(p-k)d}{p-k-i\epsilon}\right)\biggr)\,,\\
  \tilde{\psi}_{E}^{-}(p) &:=& \int_{-\infty}^{\infty}dx\,e^{-ipx}\psi_{E}^{-}(x) = \frac{1}{\sqrt{4\pi}}\biggl(2\pi\delta(p-k)-2\pi\delta(p+k) \\
   &-& ie^{ikd}\left(\frac{\tau(E)-\rho(E)-1}{\sin kd}\right)\left(\frac{e^{ipd}\sin kd - \sin(p+k)d}{p+k+i\epsilon} + \frac{e^{-ipd}\sin kd + \sin(p-k)d}{p-k-i\epsilon}\right)\biggr)\,.
\end{eqnarray*}
They are normalized as
\begin{equation}
\int_{-\infty}^{\infty}\frac{dp}{2\pi}\tilde{\psi}_{E_{1}}^{\pm\ast}(p)\tilde{\psi}_{E_{2}}^{\pm}(p) = \delta(k_{1}-k_{2})\,,
\end{equation}
where we have dropped terms $\sim\delta(k_{1}+k_{2})$ resulting from the integrals, since we label our states only by positive $k_{i}=+\sqrt{2\mu E_{i}\,}$. The above wave functions solve the Schr\"odinger equations
\begin{equation}\label{eq:Schrodingermixedform2}
  \left(\frac{p^2}{2\mu}-E\right)\tilde{\psi}_{E}^{\pm}(p) + \int_{-\infty}^{\infty}dx\,e^{-ipx}V(x)\psi_{E}^{\pm}(x)=0\,,
\end{equation}
which can be checked employing the identities $ik(\tau+\rho-1)=\mu v\cos kd(e^{-ikd}+e^{ikd}(\tau+\rho))$ and $ik(\tau-\rho-1)=i\mu v\sin kd(e^{-ikd}-e^{ikd}(\tau-\rho))$.\\
For $v<0$, there are also bound states, with energy $E_{B}<0$. Bound state wave functions of even ($+$) and odd ($-$) parity can be written as 
\begin{eqnarray}
  \psi^{+}_{B}(x) &=& \mathcal{N}^{+}_{B}\left(\theta(-d-x)e^{\kappa x} + \theta(d-x)\theta(x+d)e^{-\kappa d}\left(\frac{\cosh\kappa x}{\cosh\kappa d}\right) + \theta(x-d)e^{-\kappa x}\right)\,,\label{eq:psiBplus}\\
  \psi^{-}_{B}(x) &=& \mathcal{N}^{-}_{B}\left(\theta(-d-x)e^{\kappa x} - \theta(d-x)\theta(x+d)e^{-\kappa d}\left(\frac{\sinh\kappa x}{\sinh\kappa d}\right) - \theta(x-d)e^{-\kappa x}\right)\,,\label{eq:psiBminus}\\
  \mathcal{N}^{+}_{B} &=& \frac{\sqrt{2\kappa}\cosh\kappa d}{\sqrt{1+(1+2\kappa d)e^{-2\kappa d}}}\,,\quad \mathcal{N}^{-}_{B} = \frac{\sqrt{2\kappa}\sinh\kappa d}{\sqrt{1-(1+2\kappa d)e^{-2\kappa d}}}\,,\quad \kappa :=\sqrt{-2\mu E_{B}} >0\,,\,\, d>0\,.\nonumber
\end{eqnarray}
The corresponding Fourier transforms are denoted as $\tilde{\psi}^{\pm}_{B}(p)$. They satisfy the Schr\"odinger equations
\begin{eqnarray}
  \left(\frac{p^2}{2\mu}-E_{B}\right)\tilde{\psi}_{B}^{+}(p) &=& -\int_{-\infty}^{\infty}dx\,e^{-ipx}V(x)\psi_{B}^{+}(x) = -v\mathcal{N}^{+}_{B}e^{-\kappa d}\cos pd\,,\label{eq:SchrBplus}\\
  \left(\frac{p^2}{2\mu}-E_{B}\right)\tilde{\psi}_{B}^{-}(p) &=& -\int_{-\infty}^{\infty}dx\,e^{-ipx}V(x)\psi_{B}^{-}(x) = -v\mathcal{N}^{-}_{B}e^{-\kappa d}i\sin pd\,.\label{eq:SchrBminus}
\end{eqnarray}
In the above, $\kappa$ is a positive solution of $\,\kappa^2+\kappa\mu v +\frac{1}{4}\mu^2v^2(1-e^{-4\kappa d})\overset{!}{=}0\,$ for $v<0$. For $|\mu vd|<1$, there is just one positive solution (of ``$+$'' type\,/\,s-wave), while there are two for $|\mu vd|>1$ (an s-wave and a p-wave state, with $\kappa_{\mathrm{s}}>\kappa_{\mathrm{p}}$, solutions of $\kappa=\frac{\mu|v|}{2}(1\pm e^{-2\kappa d})\,$). \\
\begin{figure}[h]
\centering
\includegraphics[width=0.50\textwidth]{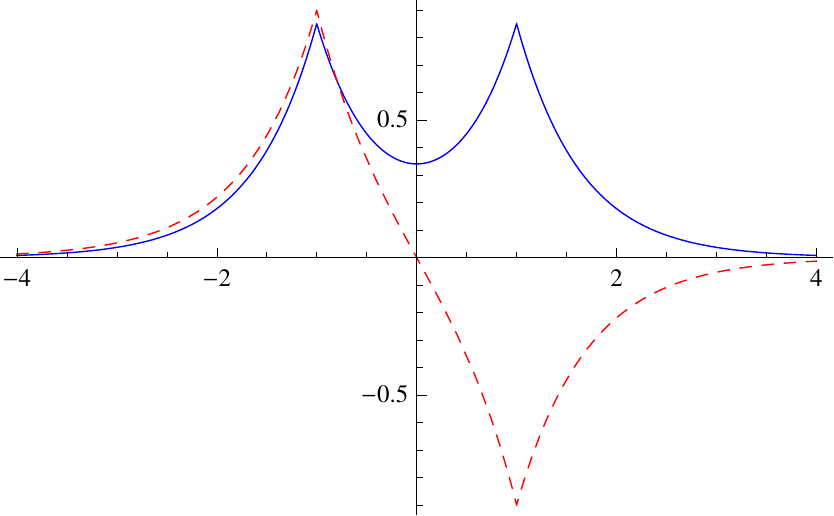}
\caption{The bound-state wave functions $\psi^{+}_{B}(x)$ (blue) and $\psi^{-}_{B}(x)$ (red, dashed) for $\mu=d=1$, $v=-3$.}
\label{fig:psiBpm}
\end{figure}%
For the case of the {\em energy-dependent\,} potential of Eq.~(\ref{eq:vsE}), the form of the solutions of the Schr\"odinger equation can be found from $\psi^{+}_{B}$, $\psi^{+}_{E}$ given above, letting $d\rightarrow 0$ there. The normalization factors might be different, however. So let us put
\begin{eqnarray*}
  \Phi_{B}(x) &=& \mathcal{N}_{B}^{\mathrm{s}}e^{-\kappa_{B}|x|}\,,\\
  \Phi_{E}(x) &=& \frac{\mathcal{N}_{E}^{\mathrm{s}}}{\sqrt{4\pi}}\left(\theta(-x)\left(e^{ikx}+(\rho_{0}(E)+\tau_{0}(E))e^{-ikx}\right) + \theta(x)\left(e^{-ikx}+(\rho_{0}(E)+\tau_{0}(E))e^{ikx}\right)\right)\,,
\end{eqnarray*}
where
\begin{equation}
\rho_{0}(E) = -\frac{i\mu v_{\mathrm{s}}(E)}{k+i\mu v_{\mathrm{s}}(E)}\,,\quad \tau_{0}(E) = \frac{k}{k+i\mu v_{\mathrm{s}}(E)}=1+\rho_{0}(E)\,.
\end{equation}
A comparison with the residue of the bound-state pole of $\mathcal{T}_{\mathrm{s}}$ shows that $\mathcal{N}_{B}^{\mathrm{s}}=\mathcal{N}_{B}^{+}\,$. Moreover, we compute (for $E_{1}\not=E_{2}$, $k_{i}=+\sqrt{2\mu E_{i}}$)
\begin{equation}
\int_{-\infty}^{\infty}dx\,\Phi^{\ast}_{E_{1}}(x)\Phi_{E_{2}}(x) = \left|\mathcal{N}_{E}^{\mathrm{s}}\right|^2\left(\delta(k_{1}-k_{2})+\frac{\tau_{0}^{\ast}(E_{1})\tau_{0}(E_{2})}{\pi}\left(\frac{v_{\mathrm{s}}(E_{1})-v_{\mathrm{s}}(E_{2})}{E_{1}-E_{2}}\right)\right)\,,
\end{equation}
which amounts to $\langle\Phi_{E_{1}}\,,\,\Phi_{E_{2}}\rangle_{V'}=\delta(k_{1}-k_{2})$ (see Eq.~(\ref{eq:newsp})) when setting $\mathcal{N}_{E}^{\mathrm{s}}=1\,$. We also confirm that $\langle\Phi_{B}\,,\,\Phi_{E}\rangle_{V'}=0$.

\section{Relation to relativistic loop integrals}
\label{app:relloops}
\def\theequation{\Alph{section}.\arabic{equation}}
\setcounter{equation}{0}
The relativistic scalar ``bubble'' integral in $d=1+1$ dimensions is
\begin{equation}
I_{\phi\phi}^{d=2}(p^2\equiv s) := \int\frac{d^2l}{(2\pi)^2}\frac{i}{((p-l)^2-m^2)(l^2-m^2)}\,,
\end{equation}
where $p=(p^{0},p^{1})$, $p^2\equiv (p^{0})^2-(p^{1})^2$, etc. The integral can directly be evaluated,
\begin{eqnarray}
  I_{\phi\phi}^{d=2}(s) &=& -\frac{1}{\pi}\int_{4m^2}^{\infty}\,\frac{ds'}{2\sqrt{s'(s'-4m^2)}(s'-s)} \nonumber \\
   &=& -\frac{1}{\pi s\sigma(s)}\mathrm{artanh}\left(-\frac{1}{\sigma(s)}\right)\,,\qquad \sigma(s)=\sqrt{1-\frac{4m^2}{s}}\,.
\end{eqnarray}
Expanding this around the threshold ($s_{thr}=4m^2$), defining $k_{cms}:=\frac{1}{2}\sqrt{s-4m^2}$, we find
\begin{equation}
I_{\phi\phi}^{d=2}(s) = -\frac{i}{8mk_{cms}}+\mathcal{O}(1)\,.
\end{equation}
We define the energy $E$ by $s=(2m+E)^2$, so that for $|E|\ll m$ we have $k_{cms}\approx\sqrt{mE}=\sqrt{2\mu E}=:k(E)\equiv k$, employing the reduced mass $\mu=m/2$. And so, in the nonrelativistic limit,
\begin{equation}\label{eq:iphiphinonrel}
-(2\pi)(2m)^2I_{\phi\phi}^{d=2}(s) \rightarrow \frac{2\pi i\mu}{k} = \int_{-\infty}^{\infty}\frac{dl}{\frac{l^2}{2\mu}-E-i\epsilon}\,.
\end{equation}
The ``triangle'' integral
\begin{eqnarray}
  I_{\Delta}^{d=2}(s,\xi^2\equiv t) &:=& \int\frac{d^{2}l}{(2\pi)^{2}}\frac{i}{((\xi-l)^2-m^{2})(l^2-m^{2})((p+l)^2-m^{2})}\biggr|_{(p+\xi)^2\overset{!}{=}p^2=s} \nonumber \\
  &=& \frac{1}{4\pi}\int_{0}^{1}dy\int_{0}^{1-y}dx\,\frac{1}{\left(sy^2-sy+m^2-tx(1-x-y)\right)^2}\label{eq:IDelta}
\end{eqnarray}
can not be given in terms of elementary functions. We find the following representation most useful:
\begin{eqnarray}
I_{\Delta}^{d=2}(s,t) &=& \frac{\theta(\mathrm{Re}\,s -2m^2)}{t_{0}(s)-t}\left(\frac{\sqrt{s-2m^2+i\sqrt{s(4m^2-s)}}}{2m^2\sqrt{2(4m^2-s)}}+\frac{\sqrt{s-2m^2-i\sqrt{s(4m^2-s)}}}{2m^2\sqrt{2(4m^2-s)}}\right) \nonumber \\ &+& \frac{1}{\pi}\int_{4m^{2}}^{\infty}dt'\,\frac{t'-2s}{4m^2\left(t'-t_{0}(s)\right)\sqrt{t'(t'-4m^2)}(t'-t)}\,,\label{eq:d2dispextension}
\end{eqnarray}
where $t_{0}(s)=4s-\frac{s^2}{m^2}\,$. The second term is the ``normal'' dispersive integral which is obtained by ``cutting'' the lines of the Feynman diagram with loop momenta $l$ and $\xi-l$ ($t$-channel cut). The first term, which only contributes when $\mathrm{Re}\,s>2m^2$, is due to the ``anomalous threshold'' singularity, stemming from the kinematic region where all three propagators in the loop can be on their mass shell. To obtain this representation, one first considers the Feynman parameter form in the second line of Eq.~(\ref{eq:IDelta}), where the integration runs over the triangle $0\leq x,\,0\leq y,\,x+y\leq 1$. To arrive at a dispersive form containing a denominator $\sim(t'-t)$, this form suggests to introduce the new variable
\begin{equation}
t'(x,y)=\frac{sy^2-sy+m^2}{x(1-x-y)}\,,
\end{equation}
which we can use to replace the Feynman parameter $x$. To do this, one has to study carefully the behavior of the function $t'(x,y)$ in the $(x,y)$ triangle of integration.
Let us, for the moment, treat $s$ as a real variable. We find that the minimum values of $t'$ (for given $y$) are lying on the line $y=1-2x$. For $s<2m^2$, the absolute minimum lies outside the triangle of integration, and the minimum value inside the triangle is $t'=4m^2$, attained at $y=0,\,x=\frac{1}{2}$. One observes that the integrand is symmetric under $x=\frac{1-y}{2}+\delta x$ $\rightarrow$ $\frac{1-y}{2}-\delta x$, and so we can split the triangle of integration along the line $y=1-2x$ and write
\begin{eqnarray*}
  \frac{1}{2}I_{\Delta}^{d=2}(s,t) &=& \frac{1}{4\pi}\int_{0}^{\frac{1}{2}}dx\int_{0}^{1-2x}dy\,\frac{1}{\left(sy^2-sy+m^2-tx(1-x-y)\right)^2}\\
  &=& \frac{1}{4\pi}\int_{0}^{\frac{1}{2}}dx\int_{0}^{1-2x}dy\,\frac{t'^{2}(x,y)}{(sy^2-sy+m^2)^2(t'(x,y)-t)^2}\,.
\end{eqnarray*}
We would now like to trade the parameter $x$ for $t'$. For a fixed value of $t'$ $\in$ $\lbrack 4m^2,\,\infty\rbrack$, $y$ assumes values between zero and $y_{\mathrm{max}}\,$ in the lower part of the split triangle,
\begin{equation}
  y_{\mathrm{max}} = 1-\frac{2\left(\sqrt{s^2+m^2(t'-4s)}-s\right)}{t'-4s}\,.
\end{equation}
So, for $s<2m^2$, we can write the integral as
\begin{equation}
I_{\Delta}^{d=2}(s,t) = \frac{1}{2\pi}\int_{4m^2}^{\infty}dt'\int_{0}^{y_{\mathrm{max}}}dy\,\left|\frac{dx}{dt'}(t',y)\right|\frac{t'^{2}}{(sy^2-sy+m^2)^2(t'-t)^2}\,,\quad \mathrm{with}
\end{equation}
\begin{equation}
\frac{dx}{dt'}(t',y) = -\frac{sy^2-sy+m^2}{t'^{\frac{3}{2}}\sqrt{t'(1-y)^2-4(sy^2-sy+m^2)}}\,.
\end{equation}
The $y$-integral can be done analytically, and after a further partial integration in $t'$ one arrives at the dispersive integral given in Eq.~(\ref{eq:d2dispextension}). Considering {\em complex\,} $s$, one notes that the integrand in this representation has a pole in the interval of $t'$-integration for $\mathrm{Re}\,s=2m^2$, so the dispersive part of Eq.~(\ref{eq:d2dispextension}) can not be analytically continued to $\mathrm{Re}\,s>2m^2$.\\
For $2m^2<s<4m^2$, there is an elliptic region contained inside the $(x,y)$ triangle of integration which is not covered by the curves for $t'>4m^2$. The function $t'(x,y)$ assumes values between $t_{0}(s)=4s-\frac{s^2}{m^2}$ and $4m^2$ there, while $y$ varies between
\begin{displaymath}
  y_{\mathrm{min}}=1-\frac{2\left(\sqrt{s^2+m^2(t'-4s)}+s\right)}{4s-t'}\quad\mathrm{and}\quad y_{\mathrm{max}}=1+\frac{2\left(\sqrt{s^2+m^2(t'-4s)}-s\right)}{4s-t'}\,.
\end{displaymath}
Note that $y_{\mathrm{min}}\rightarrow 0$ from above as $t'\rightarrow 4m^2$. So we have to consider the additional portion
\begin{equation*}
I_{\Delta,\mathrm{ano}}^{d=2} = \frac{1}{2\pi}\int_{t_{0}(s)}^{4m^2}dt'\int_{y_{\mathrm{min}}}^{y_{\mathrm{max}}}\frac{\sqrt{t'}dy}{(sy^2-sy+m^2)\sqrt{t'(1-y)^2-4(sy^2-sy+m^2)}(t'-t)^2}\,.
\end{equation*}
Moreover, for $s>2m^2$, there is a boundary term in the partial integration in $t'$ performed in the ``dispersive'' part. Adding all intermediate results up, we finally arrive at the desired representation of Eq.~(\ref{eq:d2dispextension}). The term in the first line of Eq.~(\ref{eq:d2dispextension}) tends to $(2m^2(4m^2-t))^{-1}$ if $s\rightarrow 2m^2$ from above. It cancels exactly the discontinuity of the dispersive integral in the second line at $\mathrm{Re}\,s = 2m^2$.
Consequently, this form can be analytically continued to the whole complex $s$-plane. The analytic properties in $t$ are obvious from the explicit form of the two parts of the representation.\\
One expects that the non-relativistic limit of the triangle integral will be dominated by the kinematic region where all propagators are on their mass shell, so that the first (``anomalous threshold'') term will be the most important part. In fact, the dispersive integral in the second line of Eq.~(\ref{eq:d2dispextension}) shows no singularity at $s=4m^2$, while the first term does. Keeping only the lowest-order terms in energy $E/m$, we have, in the non-relativistic limit, $t_{0}(s)\approx -16mE=-16k^2$, and 
\begin{eqnarray}
  I_{\Delta}^{d=2}(s,t) &\rightarrow& \frac{i}{2mk_{cms}(-16k_{cms}^2-t)}\,,\qquad\mathrm{or}\nonumber \\
  (2\pi)(2m)^3I_{\Delta}^{d=2}(s,t) &\rightarrow& \frac{8\pi i\mu^2}{k(Q^2-4k^2)}\,,
\end{eqnarray}
identifying $k_{cms}\rightarrow k$ and $t\rightarrow -4Q^2$. The latter replacement is not immediately intuitive. One has to realize here that (\ref{eq:IDelta}) is lorentz-invariant, while (\ref{eq:F0}) is not, and that the frame with $\xi^{0}=0$ ($\xi=(0,\xi^{1}\not=0)$) cannot be identical with the rest frame of the incoming or outgoing particle under the given kinematical conditions. Thus, one should see the above replacement as giving an invariant physical meaning to the variable $Q^2$.
\end{appendix}

\end{document}